\documentstyle[preprint,eqsecnum,aps]{revtex}
\tightenlines
\begin{document}
\title{Extension of the bilinear formalism to 
supersymmetric KdV-type equations}
\author{A. S. Carstea{\footnote {E-mail: acarst@theor1.ifa.ro}}}
\address{Institute of Physics and Nuclear 
Engineering, Dept. of Theoretical Physics, MG-6, Bucharest, Romania}

\maketitle

\begin{abstract}
Extending the gauge-invariance principle for $\tau$ functions of 
the standard bilinear formalism to the supersymmetric case,
we define N=1 supersymmetric Hirota operators.  
Using them, we bilinearize SUSY KdV-type 
equations (KdV, Sawada-Kotera-Ramani, Hirota-Satsuma). 
The solutions for multiple collisions of 
super-solitons and extension to SUSY sine-Gordon are 
also discussed.
\end{abstract}

\section{Introduction}
Supersymmetric integrable systems constitute a very interesting subject
and as a consequence a number of well known integrable equations 
have been generalized into supersymmetric (SUSY) context.
We mention the SUSY 
versions of sine-Gordon \cite{di vecchia}, \cite{chaichain},
nonlinear Schr\"odinger \cite{roelofs}, KP-hierarchy \cite{manin}, 
KdV \cite{manin},\cite{mathieu}, Boussinesq \cite{yung} etc.
We also point out that there are 
many generalizations related to the number $N$
of fermionic independent variables. In this paper we are dealing with the
$N=1$ SUSY.

So far, many of the tools used in standard theory have been extended to this
framework, such as B\"acklund transformations \cite{chaichain}, prolongation
theory \cite{roelofs}, hamiltonian formalism \cite{oevel}, 
grasmmannian description \cite{ueno}, $\tau$ functions \cite{medina}, Darboux
transformations \cite{liu}. The physical interest in the study of these systems
have been launched by the seminal paper of Alvarez-Gaume et. al \cite{alvarez}
about the partition function and 
super-Virasoro constraints of 2D quantum supergravity.
Although the $\tau$ function theory in the context of SUSY pseudodifferential
operators was given for the SUSY KP-hierarchy \cite{ueno}, 
the bilinear formalism 
for SUSY equations was very little investigated.
We mention here the algebraic approach using the
representation theory of affine Lie super-algebras in the papers of 
Kac and van der Leur \cite{kac}, Kac and Medina\cite{kacmed}
the super-conformal field theoretic approach of LeClair \cite{leclair}. 
Anyway in these articles the bilinear hierarchies are not related
to the SUSY hierarchies of nonlinear equations.

In this paper we consider a 
direct approach to SUSY equations rather than hierarchies 
namely extending the
gauge-invariance principle of $\tau$ functions for classical Hirota
operators. Our result generalize the Grammaticos-Ramani-Hietarinta
\cite{grammaticos} theorem, to SUSY case and we find 
N=1 SUSY Hirota bilinear operators. With these operators one can
obtain SUSY-bilinear forms for SUSY KdV equation of Mathieu \cite{mathieu}
and also
it allows bilinear forms for certain SUSY extensions of Sawada-Kotera-Ramani 
\cite{sw}
and Hirota-Satsuma (long water wave) \cite{hirota1976} equations.
Also the gauge-invarince principle allows to study the SUSY 
multisoliton solutions as exponentials of linears.
We want to emphasize that a special super-bilinear identity for N=1 SUSY
KdV hierarchy was conjectured by McArthur and Yung \cite{mcarthur}. Using it
they were able to write the SUSY KdV hierarchy in the bilinear form. Our
approach generalizes the super bilinear operator conjectured by them and
in the case of SUSY KdV-type equations we obtain the same results.

The paper is organized as follows. In section II the standard
bilinear formalism is briefly discussed. 
In section III supersymmetric versions
for nonlinear evolution equations are presented and in section IV 
we introduce the super-bilinear formalism.
In the last section we shall present the bilinear form for SUSY KdV-type
equations, super-soliton solutions and several comments about extension
to N=2 SUSY sine-Gordon equation.

\section{Standard bilinear formalism}

The Hirota bilinear operators were introduced as an antisymmetric extension
of the usual derivative \cite{hir}, because of their usefulness for the 
computation of multisoliton solution of nonlinear evolution equations. 
The bilinear operator 
${\bf D}_{x}=\partial_{x_{1}}-\partial_{x_{2}},$
acts on a pair of functions (the so-called "dot product") antisymmetrically:
\begin{equation}
{\bf D}_{x}f\bullet g
=(\partial_{x_{1}}-\partial_{x_{2}})f(x_{1})f(x_{2})|_{x_{1}=x_{2}=x}=f'g-fg'.
\end{equation}
The Hirota bilinear formalism has been instrumental in the derivation
of the multisoliton solutions of (integrable) nonlinear evolution equations.
The first step in the application is a dependent variable transformation
which converts the nonlinear equation into a quadratic form. This
quadratic form turns out to have the same structure as the dispersion relation
of the linearized nonlinear equation, 
although there is no deep reason for that.
This is best understood if we consider an exemple. Starting from paradigmatic
KdV equation
\begin{equation}
u_{t}+6uu_{x}+u_{xxx}=0,
\end{equation}
we introduce the substitution $u=2\partial_{x}^2\log F$ and obtain after
one integration:
\begin{equation}
F_{xt}F-F_{x}F_{t}+F_{xxxx}F-4F_{xxx}F_{x}+3F_{xx}^2=0,
\end{equation}
which can be written in the following condensed form:
\begin{equation}
({\bf D}_{x}{\bf D}_{t}+{\bf D}_{x}^4)F\bullet F=0.
\end{equation}
The power of the bilinear formalism lies in the fact that for multisoliton 
solution $F$'s are polynomials of exponentials. Moreover it displays also
the interaction (phase-shifts) 
between solitons. In the case
of KdV equation the multisoliton solution has the following form:
\begin{equation}
F=\sum_{\mu=0,1}\exp{(\sum_{i=1}^{N}\mu_{i}
\eta_{i}+\sum_{i<j}A_{ij}\mu_{i}\mu_{j})},
\end{equation}
where $\eta_{i}=k_{i}x-k_{i}^3 t+\eta_{i}^{(0)}$ and $exp{A_{ij}}=
(\frac{k_{i}-k_{j}}{k_{i}+k_{j}})^2 $ which is the phase-shift from
the interaction of the soliton $"i"$ with the soliton $"j"$.

A very important observation (which motivated the present paper)
is the relation of the physical field $u=2\partial_{x}^2\log{F}$ of KdV
equation with the Hirota function $F$: the gauge-transformation 
$F\rightarrow e^{px+\omega t}F$ leaves $u$ invariant. This is a general 
property of all bilinear equation. Moreover, one can define the Hirota
operators using the requirement of gauge-invariance. 
Let's introduce a general
bilinear expression, 
\begin{equation}
A_{N}(f,g)=\sum_{i=0}^{N}c_{i}(\partial_{x}^{N-i} f)(\partial_{x}^{i} g)
\end{equation}
and ask to be invariant under the gauge-trasformation:
\begin{equation}
A_{N}(e^{\theta}f,e^{\theta}g)=e^{2\theta}A_{N}(f,g)
\quad \theta=kx+\omega t+...(linears).
\end{equation}
Then we have the following,\cite{grammaticos}

{\it Theorem: $A_{N}(f,g)$ is gauge-invariant if and only if} 
$A_{N}(f,g)={\bf D}_{x}^{N}f\bullet g$ i.e. 
$$c_{i}=c_{0}(-1)^{i}
\left(  \begin{array}{c}
        N\\i
        \end{array}     \right)$$
and $c_{0}$ is a constant and the brakets represent binomial coefficient.

\section{Supersymmetry}

The supersymmetric extension of a nonlinear 
evolution equation (KdV for instance) refers to a system
of coupled equations for a bosonic $u(t,x)$ and a 
fermionic field $\xi(t,x)$ which reduces to the
initial equation in the limit 
where the fermionic field is zero (bosonic limit).
In the classical context, a fermionic field is described by an anticommuting
function with values in an {\it infinitely} generated Grassmann algebra.
However, supersymmetry is not just a coupling of a bosonic field to a 
fermionic field. It implies a transformation (supersymmetry invariance)
relating these two fields
which leaves the system invariant.
In order to have a mathematical formulation of these concepts we have to 
extend the classical space $(x,t)$ to a larger space (superspace) 
$(t,x,\theta)$  
where $\theta$ is a Grassmann variable and 
also to extend the pair of fiels $(u,\xi)$ to a larger
fermionic or bosonic superfield $\Phi(t,x,\theta).$ 
In order to have nontrivial extension for KdV we choose $\Phi$ to 
be fermionic, having the expansion
\begin{equation}
\Phi(t,x,\theta)=\xi(t,x)+\theta u(t,x).
\end{equation}
The N=1 SUSY means that we have
only one Grassmann variable $\theta$ and 
we consider only space supersymmetry invariance
namely $x\rightarrow x-\lambda\theta$ and $\theta\rightarrow \theta+\lambda$
($\lambda$ is an anticommuting parameter). This transformation is generated
by the operator 
$Q=\partial_{\theta}-\theta\partial_{x},$ 
which anticommutes
with the covariant derivative 
$D=\partial_{\theta}+\theta\partial_{x}$ (Notice also that $D^2=\partial_{x}$). 
Expressions written in terms of the covariant derivative and the superfield
$\Phi$ are manifestly supersymmetric invariant.
Using the superspace formalism one can 
construct different supersymmetric extension of 
nonlinear equations. Thus the integrable (in the sense of Lax pair) 
variant of N=1 SUSY KdV is \cite{manin}\cite{mathieu}
\begin{equation}\label{skdv}
\Phi_{t}+D^6 \Phi+3D^2(\Phi D\Phi)=0,
\end{equation}
which on the components has the form
\begin{eqnarray}
u_{t}&=&-u_{xxx}-6uu_{x}+3\xi\xi_{xx}\nonumber\\
\xi_{t}&=&-\xi_{xxx}-3\xi_{x}u-3\xi u_{x}.
\end{eqnarray}

We shall discuss also the following supersymmetric equations, although
we do not know if these equations are completely integrable in the sense
of Lax pair ($\Phi$ is also a fermionic superfield).

\begin{itemize}
\item N=1 SUSY Sawada-Kotera-Ramani,
\begin{equation}\label{sskr}
\Phi_{t}+D^{10} \Phi+D^2(10 D\Phi D^4 \Phi+5 D^5 \Phi\Phi+15(D\Phi)^2 \Phi)=0.
\end{equation}

\item N=1 SUSY Hirota-Satsuma (shallow water wave)
\begin{equation}\label{shs}
D^4 \Phi_{t}+\Phi_{t} D^3 \Phi+2D^2 \Phi D \Phi_{t}-D^2 \Phi-\Phi_{t}=0.
\end{equation}
\end{itemize}

A very important equation from the physical consideration is the
SUSY sine-Gordon. We are going to consider the version studied by
Kulish and Tsyplyaev \cite{tsy}. There are other integrable versions
of SUSY sine-Gordon emerged from algebraic procedures \cite{inami}. 
In this case one needs two Grassmann variables 
$\theta_{\alpha}$ with $\alpha=1,2$
and the supersymmetry transformation is
$$x^{'\mu}=x^{\mu}-i\bar\lambda\gamma^{\mu}\theta, \quad 
\theta_{\alpha}^{'}=\theta_{\alpha}+\lambda_{\alpha}, \mu=1,2.$$
Here, $\lambda_{\alpha}$ is the anticommuting spinor 
parameter of the transformation and 
$\bar\lambda=(\lambda^1, \lambda^2)$,
$\lambda^{\alpha}=\lambda_{\beta}(i\sigma_{2})^{\beta\alpha}$, 
$\gamma^{0}=i\sigma_{2}$, 
$\gamma^{1}=\sigma_{1}$, 
$\gamma^{5}=\gamma^{0}\gamma^{1}=\sigma_{3}$.
We use the metric $g^{\mu\nu}=diag(-1,1)$ and $\sigma_{i}$ are the Pauli 
matrices.
The superfield has the following expansion:
\begin{equation}
\Phi(x^{\mu}, \theta_{\alpha})=\phi(x^{\mu})+i\bar\theta\psi(x^{\mu})+
\frac{i}{2}\bar\theta \theta F(x^{\mu}),
\end{equation}
where $\phi$ and $F$ are real bosonic (even) scalar fields and $\psi_{\alpha}$
is a Majorana spinor field.
The SUSY sine-Gordon equation is:
\begin{equation}\label{ssg}
{\bar D} D\Phi=2i\sin{\Phi},
\end{equation}
where $D_{\alpha}=
\partial_{\theta^{\alpha}}+i(\gamma^{\mu}\theta)_{\alpha}\partial_{\mu}$
and on the components it has the form:
\begin{eqnarray}
(\gamma^{\mu}\partial_{\mu}+\cos{\phi})\psi=0\nonumber\\
\phi_{xx}-\phi_{tt}=\frac{1}{2}(\sin{(2\phi)}-i\bar\psi \psi \sin{\phi}).
\end{eqnarray} 
This version of SUSY sine-Gordon equation has been studied by Kulish
and Tsyplyaev \cite{tsy} using the Inverse Scattering Method. They
also found super-kink solutions.

\section{Super-Hirota operators}

In order to apply the bilinear formalism on these equations one 
has to define a SUSY bilinear operator. We are going to consider the
following general $N=1$ SUSY bilinear expression
\begin{equation}\label{susy}
S_{N}(f,g)=\sum_{i=0}^{N}c_{i}(D^{N-i} f)(D^{i} g),
\end{equation}
where $D$ is the covariant derivative and $f$, $g$ are Grassmann valued
functions (odd or even). We shall prove the following

{\it Theorem}: The general $N=1$ SUSY bilinear expression (\ref{susy})
is super-gauge invariant i.e. 
for $\Theta=kx+\omega t+\theta \hat \zeta+$...linears ($\zeta$ is a Grassmann 
parameter)
$$S_{N}(e^{\Theta}f,e^{\Theta}g)=e^{2\Theta}S_{N}(f,g),$$
if and only if
$$c_{i}=c_{0}(-1)^{i|f|+\frac{i(i+1)}{2}}\left[  \begin{array}{c}
                                                     N\\i
                                                 \end{array}     \right],$$
where the super-binomial coefficients are defined by:
$$\left[\begin{array}{c}        
              N\\i                        
        \end{array}\right]=                        
\left\{ \begin{array}{ll}
        \left(\begin{array}{c}        
              $[N/2]$   \\   $[i/2]$                        
        \end{array}\right) & \mbox {if $(N,i)\neq(0,1) mod 2$} \\
                         0 & \mbox {otherwise}                      
        \end{array}           
\right. $$
$|f|$ is the Grassmann parity of the function $f$ defined by:
$$|f|=\left\{ \begin{array}{ll}
              1 & \mbox {if $f$ is odd (fermionic)} \\
              0 & \mbox {if $f$ is even (bosonic)}                      
              \end{array}           
\right. $$
and $[k]$ is the integer part of the real number $k$ ($[k]\leq k<[k]+1$)

Proof: First we are going to consider $N$ even and we shall take it on the
form $N=2P$. In this case we have:
$$
S_{N}(f,g)=\sum_{i=1}^{N}c_{i}(D^{N-i} f)(D^{i} g)=
\sum_{i=0}^{P}c_{2i}(\partial^{P-i} f)(\partial^{i} g)
+\sum_{j=0}^{P-1}c_{2j+1}(\partial^{P-j-1} Df)(\partial^{j} Dg)
$$
Imposing the super-gauge invariance and expanding the covariant 
derivatives we obtain:
$$
\sum_{n\geq 0}\sum_{m\geq 0}\left(
\sum_{i=0}^{P}c_{2i}
\left(  \begin{array}{c}
        i\\n
        \end{array}     \right)
\left(  \begin{array}{c}
        P-i\\m
        \end{array} \right)
k^{P-n-m}\right)(\partial^m f)(\partial ^n g)+$$
$$
+\sum_{n'\geq 0}\sum_{m'\geq 0}\Lambda\left(
\sum_{j=0}^{P-1}c_{2j+1}
\left(  \begin{array}{c}
        j\\n'
        \end{array}     \right)
\left(  \begin{array}{c}
        P-j-1\\m'
        \end{array} \right)
k^{P-n'-m'-1}\right)(\partial^{m'} f)(\partial ^{n'} Dg)+$$
$$
+\sum_{n'\geq 0}\sum_{m'\geq 0}\Lambda(-1)^{|f|+1}\left(
\sum_{j=0}^{P-1}c_{2j+1}
\left(  \begin{array}{c}
        j\\n'
        \end{array}     \right)
\left(  \begin{array}{c}
        P-j-1\\m'
        \end{array} \right)
k^{P-n'-m'-1}\right)(\partial^{m'} Df)(\partial ^{n'} g)+$$
$$
+\sum_{n\geq 0}\sum_{m\geq 0}\left(
\sum_{j=0}^{P-1}c_{2j+1}
\left(  \begin{array}{c}
        j\\n
        \end{array}     \right)
\left(  \begin{array}{c}
        P-j-1\\m
        \end{array} \right)
k^{P-n-m-1}\right)(\partial^m Df)(\partial ^n Dg)=$$
$$
=\sum_{i=0}^{P}c_{2i}(\partial^{P-i} f)(\partial^{i} g)
+\sum_{j=0}^{P-1}c_{2j-1}(\partial^{P-j-1} Df)(\partial^{j} Dg)
$$
where $\Lambda=\hat\zeta+\theta k$. From this, we must have for 
every $m$, $n$ subjected to $0\leq n\leq i\leq P-m$
and $j\leq P-m'$. 
\begin{equation}\label{ddiscret}
\sum_{i=0}^{P}c_{2i}
\left(  \begin{array}{c}
        i\\n
        \end{array}     \right)
\left(  \begin{array}{c}
        P-i\\m
        \end{array} \right)
k^{P-n-m}=c_{2n}\delta_{P-n-m}
\end{equation}
Also due to the fact that the supergauge invariance has to be obeyed
for every $f$ and $g$ we must have $c_{2j+1}=0$ 
The discrete equation (\ref{ddiscret}) was solved in \cite{grammaticos}.
Its general solution is given by:
\begin{eqnarray}\label{par}
c_{2i} &=& c_{0}(-1)^{i}
\left(  \begin{array}{c}
        P\\i
        \end{array}     \right)\nonumber\\
c_{2j+1} &=& 0
\end{eqnarray}

In the case of $N=2P+1$ we proceed in a similar manner and we obtain 
the following system:
\begin{equation}\label{discret}
\sum_{i=0}^{P}c_{2i}
\left(  \begin{array}{c}
        i\\n
        \end{array}     \right)
\left(  \begin{array}{c}
        P-i\\m
        \end{array} \right)
k^{P-n-m}=c_{2n}\delta_{P-n-m}
\end{equation}
\begin{equation}\label{discret}
\sum_{j=0}^{P}c_{2j+1}
\left(  \begin{array}{c}
        j\\n
        \end{array}     \right)
\left(  \begin{array}{c}
        P-j-1\\m
        \end{array} \right)
k^{P-n-m-1}=c_{2n+1}\delta_{P-n-m-1}
\end{equation}
\begin{equation}
(-1)^{|f|}c_{2i}+c_{2i+1}=0
\end{equation}
This system has the following solution:
\begin{eqnarray}\label{impar}
c_{2i} &=& c_{0}(-1)^{i}
\left(  \begin{array}{c}
        P\\i
        \end{array}     \right)\nonumber\\
c_{2i+1} &=& c_{0}(-1)^{i+1+|f|}
\left(  \begin{array}{c}
        P\\i
        \end{array}     \right)
\end{eqnarray}
The relations (\ref{par}), (\ref{impar}) can be written in a compact form
as 
$$c_{i}=c_{0}(-1)^{i|f|+\frac{i(i+1)}{2}}\left[  \begin{array}{c}
                                                     N\\i
                                                 \end{array}     \right].$$
and the theorem is proved.  We mention that the super-bilinear operator
proposed by McArthur and Yung \cite{mcarthur} 
is a particular case of the above super-Hirota
operator.

We shall note the bilinear operator as
$$S_{N}(f,g):={\bf S}_{x}^{N}f\bullet g$$
Also, one can easily obtain the following properties:
\begin{equation}
{\bf S}_{x}^{2N}f\bullet g={\bf D}_{x}^{N}f\bullet g
\end{equation}
\begin{equation}
{\bf S}_{x}^{2N+1}e^{\eta_{1}}\bullet e^{\eta_{2}}=
[\hat\zeta_{1}-\hat\zeta_{2}+\theta(k_{1}-k_{2})](k_{1}-k_{2})^{N}
e^{\eta_{1}+\eta_{2}}
\end{equation}
\begin{equation}
{\bf S}_{x}^{2N+1}1\bullet e^{\eta_{1}}
=(-1)^{N+1}(\hat\zeta+\theta k)k^{N}e^{\eta}=
(-1)^{N+1}{\bf S}_{x}^{2N+1}e^{\eta}\bullet 1
\end{equation}
where $\eta_{i}=k_{i}x+\theta\hat\zeta_{i}$ and 
$\hat\zeta_{i}$ are odd Grassmann
numbers.

\section{Bilinear SUSY KdV-type equations}

In order to use the super-bilinear operators defined above we shall consider
the following nonlinear substitution for the superfield:
\begin{equation}
\Phi(t,x,\theta)=2D^{3}\log{\tau(t,x,\theta)}
\end{equation}
Introducing in SUSY KdV (\ref{skdv}) we obtain the following 
super-bilinear form:
\begin{equation}
({\bf S}_{x}{\bf D}_{t}+{\bf S}_{x}^{7})\tau\bullet \tau=0,
\end{equation}
which is equivalent with  the form found by McArthur and Yung\cite{mcarthur}
\begin{equation}
{\bf S}_{x}({\bf D}_{t}+{\bf D}_{x}^{3})\tau\bullet \tau=0.
\end{equation}
The 1 super-soliton solution has the following structure
\begin{equation}
\tau^{(1)}=1+e^{kx-k^3 t+\theta\hat\zeta+\eta^{(0)}}
\end{equation}
In order to find 2 super-soliton solution we are going to consider the form
\begin{equation}
\tau^{(2)}=1+e^{\eta_{1}}+e^{\eta_{2}}+
e^{\eta_{1}+\eta_{2}+A_{12}}
\end{equation}
and we have to find the factor $\exp{A_{12}}$, 
where 
$\eta_{i}=k_{i}x-k_{i}^{3} t+\theta\hat\zeta_{i}+\eta_{i}^{(0)}$
The equation for $\exp{A_{12}}$ is the following:
\begin{equation}\label{inter}
[(\hat\zeta_{1}-\hat\zeta_{2})+\theta(k_{1}-k_{2})](k_{1}-k_{2})=
\exp{A_{12}}
[(\hat\zeta_{1}+\hat\zeta_{2})+\theta(k_{1}+k_{2})](k_{1}+k_{2})
\end{equation}
We assume that $\exp{A_{12}}$ depends only on $k_{i}$, $\hat\zeta_{i}$, with
$i=1,2$ and in the bosonic limit ($\hat\zeta_{i}=0$) to have the standard
form, $(k_{1}-k_{2})^{2}/(k_{1}+k_{2})^2$. Accordingly, in order to solve
(\ref{inter}) we consider the ansatz:
\begin{equation}\label{form}
\exp{A_{12}}=(\frac{k_{1}-k_{2}}{k_{1}+k_{2}})^{2}
+\hat a(k_{1}, k_{2})\hat\zeta_{1}+\hat b(k_{1}, k_{2})\hat\zeta_{2}
+\gamma(k_{1}, k_{2})\hat\zeta_{1}\hat\zeta_{2}
\end{equation}
where $\hat a$, $\hat b$ are odd Grassmann functions depending on $k_{1}$
and $k_{2}$ and $\gamma$ is an even Grassmann function.
Introducing (\ref{form}) in (\ref{inter}) we shall find that
$$\hat a(k_{1}, k_{2})\hat\zeta_{1}+\hat b(k_{1}, k_{2})\hat\zeta_{2}
+\gamma(k_{1}, k_{2})\hat\zeta_{1}\hat\zeta_{2}=0$$
and 
$$k_{1}\hat\zeta_{2}=k_{2}\hat\zeta_{1}$$
So, the interaction effect remains the same as in the bosonic case.
One can easily verify that the N super-soliton solution is given by
\begin{equation}
\tau^{(N)}=
\sum_{\mu=0,1}\exp{(\sum_{i=1}^{N}
\mu_{i}\eta_{i}+\sum_{i<j}A_{ij}\mu_{i}\mu_{j})},
\end{equation}
where
$$\eta_{i}=k_{i}x-k_{i}^{3} t+\theta\hat\zeta_{i}+\eta_{i}^{(0)}$$
$$\exp{A_{ij}}=\left(\frac{k_{i}-k_{j}}{k_{i}+k_{j}}\right)^{2}$$
$$k_{i}\hat\zeta_{j}=k_{j}\hat\zeta_{i}$$

For N=1 SUSY Sawada-Kotera-Ramani (\ref{sskr}) using the same nonlinear
substitution, 
$$\Phi=2D^3\log{\tau(t,x,\theta)}$$
we shall find the following super-bilinear form:
\begin{equation}
({\bf S}_{x}{\bf D}_{t}+{\bf S}_{x}^{11})\tau\bullet \tau=0
\end{equation}
In a similar way we find the N super-soliton solution
\begin{equation}
\tau^{(N)}=
\sum_{\mu=0,1}
\exp{(\sum_{i=1}^{N}\mu_{i}\eta_{i}+\sum_{i<j}A_{ij}\mu_{i}\mu_{j})},
\end{equation}
where
$$\eta_{i}=k_{i}x-k_{i}^{5} t+\theta\hat\zeta_{i}+\eta_{i}^{(0)}$$
$$\exp{A_{ij}}=\left(\frac{k_{i}-k_{j}}{k_{i}+k_{j}}\right)^{2}
\frac{k_{i}^2 -k_{i}k_{j}+k_{j}^{2}}{k_{i}^2 +k_{i}k_{j}+k_{j}^{2}}$$
$$k_{i}\hat\zeta_{j}=k_{j}\hat\zeta_{i}$$

For N=1 SUSY Hirota-Satsuma equation (\ref{shs})
using the nonlinear substitution 
$$\Phi=2D\log{\tau(t,x,\theta)}$$
one obtains the  
super-bilinear form:
\begin{equation}
({\bf S}_{x}^{5}{\bf D}_{t}-{\bf S}_{x}^{3}
-{\bf S}_{x}{\bf D}_{t})\tau\bullet \tau=0
\end{equation}
The N super-soliton solution is,
\begin{equation}
\tau^{(N)}=
\sum_{\mu=0,1}\exp{(\sum_{i=1}^{N}\mu_{i}
\eta_{i}+\sum_{i<j}A_{ij}\mu_{i}\mu_{j})},
\end{equation}
where
$$\eta_{i}=k_{i}x-k_{i} t/(k_{i}^2-1)+\theta\hat\zeta_{i}+\eta_{i}^{(0)}$$
$$\exp{A_{ij}}=
\left(\frac{k_{i}-k_{j}}{k_{i}+k_{j}}\right)^{2}
\frac
{ (k_{i}-k_{j})^2+k_{i}k_{j}[(k_{i}-k_{j})^2-(k_{i}^2 -1)(k_{j}^2 -1)]}
{ (k_{i}-k_{j})^2-k_{i}k_{j}[(k_{i}-k_{j})^2-(k_{i}^2 -1)(k_{j}^2 -1)]}
$$
$$k_{i}\hat\zeta_{j}=k_{j}\hat\zeta_{i}$$

We can ask ourselves if it is possible to obtain super-bilinear forms
for SUSY equations of the nonlinear Klein-Gordon type. In fact 
the SUSY sine-Gordon equation(\ref{ssg})
can be written in the following form:
\begin{equation}
[D_{T}, D_{X}]\Phi(T, X, \theta, \theta_{t})
=2i\sin{\Phi(T, X, \theta, \theta_{t})}
\end{equation}
where we have introduced the light-cone variables
$X:=i(t-x)/2$, $T:=i(t+x)/2$, and 
$\theta:=\theta_{1}$, $\theta_{2}:=-\theta_{t}.$
Covariant derivatives are
$D_{X}:=\partial_{\theta}+\theta\partial_{X}$, 
$D_{T}:=\partial_{\theta_{t}}+\theta_{t}\partial_{T}$
and the square braket means the commutator.
Using the nonlinear substitution ($G$ and $F$ are even functions)
$$\Phi=2i\log{\left(\frac{G}{F}\right)},$$
we find the following quadrilinear expression
$$2i\{F^2
(G[D_{T},D_{X}]G-[D_{T}G, D_{X}G])
-G^2(F[D_{T},D_{X}]F-[D_{T}F, D_{X}F])
\}=F^4-G^4$$
It is easy to see that the bilinear operator
$${\bf S}_{XT}\tau\bullet \tau:=\tau[D_{T},D_{X}]\tau-[D_{T}\tau, D_{X}\tau]$$
is super-gauge invariant with respect to the super-gauge
$$e^{\Theta}:=e^{(kx+\omega t+\theta\hat\zeta+\theta_{t}\hat\Omega+liniars)}.$$
Accordingly
we can choose the following super-bilinear form, formally the same with 
standard sine-Gordon equation,
\begin{eqnarray}
{\bf S}_{XT}G\bullet G &=& \frac{1}{2i}(F^2-G^2)\nonumber\\
{\bf S}_{XT}F\bullet F &=& \frac{1}{2i}(G^2-F^2)
\end{eqnarray}
but, it is not clear how to compute the super-kink solutions.

From these examples it seems that gauge-invariance is a useful concept
for bilinear formalism in the supersymmetric case, 
though there is no deep reason for that. 
As a consequence we was able
to bilinearize several supersymmetric equations of KdV type. The case
of SUSY versions for mKdV, NLS, KP etc. requires further 
investigation because it seems that {\it only
certain supersymmetric extensions are super-bilinearizable}.
Although we do not know if the SUSY extension of 
Sawada-Kotera and Hirota-Satsuma proposed above
are integrable in the sense of Lax, they admit super-bilinear form and
also N super-soliton solution. Accordingly, the  
integrability in the sense of Hirota is satisfied. Probably a singularity
analysis implemented on the super-bilinear form will reveal
the connection between Hirota-integrability and Lax-integrability.
\vskip 2cm
\noindent ACKNOWLEDGEMENTS: 

\noindent This work was supported by the grant Nr. B1/18 MCT, 1998
of the Romanian Ministery of Research. A part of this work was done
at the Institute of Theoretical Physics, University of Bern, Switzerland.
The author wants to express his sincere thanks to Prof. H. Leutwyler
for hospitality.

\end{document}